\documentclass[12pt,a4paper]{article}
\usepackage[textwidth=450pt, textheight=660pt]{geometry}
\usepackage{amsmath,amsthm,amssymb}
\usepackage{graphicx,tabu}
\usepackage[pagebackref=false]{hyperref}
\usepackage[active]{srcltx}
\usepackage[affil-it]{authblk}
\usepackage{wrapfig}
\usepackage[noadjust]{cite}
\usepackage{filecontents}
 \usepackage{color,soul}
  \usepackage{xcolor}
  
 \newcommand{\be}{\begin{equation}}
\newcommand{\ee}{\end{equation}}
\newcommand{\bea}{\begin{eqnarray}}
\newcommand{\eea}{\end{eqnarray}}
\newcommand{\nn}{\nonumber}

\begin{document}

  \vspace{2cm}

  \begin{center}
    \font\titlerm=cmr10 scaled\magstep4
    \font\titlei=cmmi10 scaled\magstep4
    \font\titleis=cmmi7 scaled\magstep4
  {\bf  Casimir  energy  for  acoustic phonons in Graphene }

    \vspace{1.5cm}
     \noindent{{\large Y. Koohsarian ${}^a$ \footnote{yo.koohsarian@mail.um.ac.ir}, K. Javidan ${}^a$ \footnote{Javidan@um.ac.ir},
  A. Shirzad  ${}^{b,c}$ \footnote{shirzad@ipm.ir}}} \\
     ${}^{a}$ {\it Department of Physics, Ferdowsi University of Mashhad \\
       P.O.Box  91775-1436, Mashhad, Iran} \\
   ${}^b$ {\it Department of Physics, Isfahan University of Technology \\
       P.O.Box 84156-83111, Isfahan, Iran,\\
      ${}^c$  School of Physics, Institute for Research in Fundamental Sciences (IPM),\\
       P. O. Box 19395-5531, Tehran, Iran} \\

  \end{center}
  \vskip 2em

\begin{abstract}
We find the Casimir energy, at finite temperature, for  acoustic  phonons in a   Graphene sheet suspended over a rectangular trench, and the corresponding Casimir forces are interpreted  as   correction terms   to the built-in   tensions of the Graphene. We show  that   these corrections generally break the tensional isotropy of the membrane, and can increase or decrease the membrane tension.  We demonstrate that  for a narrow rectangular trench with   side-lengths in the order of   few   nanometers and   few micrometers, these temperature corrections are expected to be  noticeable ($\sim 10^{-4} N/m$) at the room temperature.  These   corrections would  be even more considerable  by increasing the temperature, and can be applied for adjusting the   built-in tension of the Graphene. Consequently we introduce a corrected version for the fundamental resonance frequency of the Graphene resonator. 
\end{abstract}

 \textbf{Keywords} \\   Casimir energy, Acoustic  Phonons,  Membrane, Temperature correction, built-in tension, Resonance frequency, Graphene sheet.

\section{Introduction} \label{sec-1}
The Casimir effect, as one of the most important manifestations of the zero-point oscillations of quantum fields \cite {Cas}, nowadays has been investigated theoretically and experimentally for various systems with different boundary conditions and different geometries, see e.g. \cite{CR1,CR2,CR3} as  review. As we know, the Casimir forces are observable mostly in microscale and smaller dimensions, so the Casimir effect can  imply  significant   results  for  ultra-small  structures, see \cite{CNR1,CNR2,CNR3} as review. Graphene, a single-atom-thick layer of Carbon atoms covalently bounded  in a honeycomb lattice, as the thinnest known material with remarkable mechanical and electrical properties, is expected to have many promising applications in ultra-small-dimension systems, see e.g. \cite{NEMS1,NEMS2,NEMS3}.  

 In this paper, we investigate the  physical effects of the zero-point oscillations      for   acoustic phonons in a monolayer Graphene sheet suspended over a rectangular trench. Here the  acoustic phonons are considered as the quanta of a massless bosonic field with two components, longitudinal acoustic (LA) and transverse  acoustic (TA) fields, constrained by a Dirichlet boundary condition which is resulted from the strong Van der Waals (VdW) forces which clamp the Graphene sheet to the sidewalls of the trench, see Refs. \cite{BT,GV1,GV2,GV3,GV4,GT1,GT2,GT3}.  So one can introduce a  ``phononic'' Casimir  energy  for the Graphene sheet by calculating the   zero-point energy of of these constrained  acoustic fields in the suspended Graphene.  In fact here the Graphene sheet can be considered as a phonon gas  being confined in a 2-dimensional box with sidelengths equivalent to the trench sidelengths. As a result , the    corresponding  ``phononic'' Casimir forces can be defined by differentiating the above Casimir energy with respect to the side-lengths of the trench, see Sec. 2.  We would apply a new useful technique to find an exact expression for the Casimir energy being specifically useful for high temperatures, which has considerable importance for our work, since as we will see, the effective temperature of the Graphene membrane is considerably smaller than the  ordinary temperatures. Then we would interpret these  Casimir forces as temperature-dependent corrections to the built-in  tension of the membrane, and using the experimental values, we obtain numerical results for these temperature corrections, see Sec. 3. Subsequently, utilizing these temperature corrections, we introduce a temperature-dependent corrected version for the fundamental resonance frequency of the Graphene membrane.

\section{Casimir energy for  acoustic phonons  in a  membrane}

As we know, the Dynamics of the acoustic modes in a membrane  can be effectively  described by the known dispersion relations $\omega_{I}= \upsilon_{I} k $ in which $\omega$ and $k$ are the mode frequency and the mode wavenumber, respectively,  of the acoustic modes,   $\upsilon$ is the sound velocity in the membrane, and $I$ counts  the longitudinal (LA) and transverse (TA) components.  As a result for  a fully clamped  membrane,  these acoustic modes can be regarded as oscillation modes of a massless 2-component bosonic field living in a rectangle with Dirichlet boundary conditions, with the known mode wavenumbers 
\be
k_{n,m} =\pi \sqrt{\frac{n^2}{a^2} +\frac{m^2}{b^2} }; \ \ \ a\geq b, \ \ \  n,m=1,2,... \label{eq-1}
\ee
in which, ``$a$" and ``$b$" are the side-lengths of  the rectangular trench, over which, the membrane is suspended. Now one can quantize the system simply by taking the  oscillation modes  of the mentioned acoustic fields, as quantum oscillators with mode frequencies $\omega_{I,n,m}= \upsilon_{I} k_{n,m}$ given through Eq. \eqref{eq-1}. Then the zero-point energy of such a  system at finite temperature, can be written as  (see e.g. \cite{ACE})
 \be
E_0(a,b,T) = \frac{k_B T}{2}\sum_{I=\textrm{LA,TA}} \ \sum_{l=-\infty}^{\infty} \sum_{n,m=1}^{\infty}  \ln\left[\left(\frac{2\pi k_B T}{\hbar}\right)^2 l^2+ \omega_{I, n,m}^2 \right]. \label{eq-2}
 \ee
in which, ``$T$" is the system temperature, and $k_B$ and $\hbar$ are the Boltzmann and the Planck constants respectively.  Note that, as we discussed before, the above zero-point energy can be considered actually as the vacuum energy of the membrane acoustic phonons considered as a phonon gas being confined in  a 2-dimensional box with sidelengths ``$a$" and ``$b$". So the regularized form of the above zero-point energy can be considered as the ``phononic'' Casimir energy of the membrane.  

Now to find an explicit expression for the phononic  zero-point energy \eqref{eq-2}, utilizing the Gamma function,  we rewrite the above equation as a parametric integral;
\bea
E_0(a,b,T)& =& - \frac{k_B T}{2} \sum_I \sum_{l } \sum_{n,m} \lim_{s\rightarrow 0} \frac{\partial}{\partial s} \left(\frac{\hbar \nu}{ 2\pi k_B T} \right)^{2s}\left(  l^2+ \lambda_{I,a}^2 n^2 + \lambda_{I,b}^2 m^2  \right)^{-s} \nn \\
&=&  - \frac{k_B T}{2} \sum_I \sum_{l } \sum_{n,m} \lim_{s\rightarrow 0} \frac{\partial}{\partial s} \left(\frac{\hbar \nu}{ 2\pi k_B T} \right)^{2s} \nn \\
 && \hspace{3 cm} \times \int_0^{\infty} \frac{dt}{t} \frac{t^s}{\Gamma(s)} \exp \left[-t\left( l^2+ \lambda_{I,a}^2 n^2 + \lambda_{I,b}^2 m^2  \right)\right] \label{eq-3}
\eea
in which $\nu$ is an arbitrary parameter with the dimension of the inverse time, and  we have introduced dimensionless variables $\lambda = \theta/T$ with effective temperatures
\bea
  \theta_{I,a} = \frac{ \hbar \upsilon_I}{2 a k_B} , \  \ \theta_{I,b} = \frac{\hbar \upsilon_I}{2 b k_B}. \label{eq-4}
\eea
  Note that it is specifically important, for our problem, to find an explicit  expression for the zero-point energy for sufficiently high temperatures,  since, as we will see in the section 3, the (larger) effective temperature $\theta_{\textrm{LA},b}$ of  the suspended  Graphene membrane, is sufficiently  lower than  the ordinary temperatures. 
   As we know,  a limiting expression for the Casimir energy at high temperature, can be obtained generally by  applying the known \emph{ heat kernel} expansion, see e.g. \cite{ACE,math2}.  But here we use a new useful technique  to find an  exact  expression for the zero-point energy, being  specifically useful for  high temperatures.  First separating $l=0$ from the $l$-sum, we rewrite Eq. \eqref{eq-3} as
\bea
&&E_0(a,b,T)=  - \frac{k_B T}{2}\sum_{I}\lim_{s\rightarrow 0} \frac{\partial}{\partial s}  \left(\frac{\hbar \nu}{ 2\pi k_B T} \right)^{2s}  \left[\textbf{I}+\textbf{II}\right]; \nn \\
&&\textbf{I} \equiv \int_0^{\infty} \frac{dt}{t}\frac{t^s}{\Gamma(s)}  \sum_{n =1}^{\infty}  \exp \left[-t   \lambda_{I,a}^2 n^2  \right] \sum_{ m=1}^{\infty} \exp \left[-t   \lambda_{I,b}^2 m^2  \right] \nn \\
&&\textbf{II} \equiv 2 \int_0^{\infty} \frac{dt}{t} \frac{t^s}{\Gamma(s)}  \sum_{l=1}^{\infty}  \exp \left[-t l^2  \right] \sum_{n =1}^{\infty}  \exp \left[-t   \lambda_{I,a}^2 n^2  \right] \sum_{ m=1}^{\infty} \exp \left[-t   \lambda_{I,b}^2 m^2  \right]  \label{eq-5}
\eea
Then, by applying the     Poisson summation formula (see e.g. \cite{math1})
\be
\sum_{n=1}^{\infty} f(n)= -\frac{f(0)}{2} + \int_0^{\infty} f(x) dx +2 \sum_{n=1}^{\infty} \int_0^{\infty} f(x) \cos(2\pi nx)dx \label{eq-6}
\ee
to the $m$-sum;
\be
 \sum_{ m=1}^{\infty} \exp \left[-t   \lambda_{I,b}^2 m^2  \right]=-\frac{1}{2}+ \frac{1}{2 \lambda_{I,b}} \sqrt{\frac{\pi}{t}}+ \frac{1}{2 \lambda_{I,b}}\sqrt{\frac{\pi}{t}} \sum_{m=1}^{\infty} \exp{\left[-\frac{\pi^2 m^2}{ t\lambda_{I,b}} \right]}. \label{eq-7}
\ee
and substituting it in the expression \textbf{I} in Eq. \eqref{eq-5}, and after some calculations we find
\bea
&&\textbf{I}= -\frac{\zeta (2 s)}{2\lambda _{I,a}^{2 s} } +\frac{\sqrt{\pi } }{ 2 \lambda _{I,b}} \frac{ \Gamma \left(s-1/2\right)}{ \Gamma (s)}\frac{ \zeta (2 s-1)}{\lambda _{I,a}^{ 2 s-1}} \nn \\ 
&& \hspace{3 cm} +2 \sqrt{\pi}{\lambda _{I,b} \Gamma(s)}  \sum_{n,m=1}^{\infty} \left( \frac{n \lambda_{I,a}}{m \pi/\lambda_{I,b}}\right)^{ -s+1/2} K_{ -s+1/2}\left( 2 \pi  n m\frac{ \lambda _{I,a}}{\lambda _{I,b}} \right)
\label{eq-8}
\eea
in which, $K$ is a Bessel function of the second kind, and we have used the known Riemann zeta function $ \zeta (s)=\sum_{n=1}^{\infty} n^{-s}$, and the integral relation
\be
\int_0^{\infty}  t^r \exp\left[-x^2 t-y^2/t \right]dt=2(x/y)^{-r-1}K_{-r-1}(2xy). \label{eq-9}
\ee
Then, for the expression \textbf{II}, applying the   Poisson summation formula \eqref{eq-6} to both $m$-  and $n$-sum, and after some  similar calculations, one can find

\bea
&&\textbf{II}=\frac{\zeta (2 s)}{2}-\frac{\sqrt{\pi }}{2} \left(\frac{1}{\lambda _{I,b}}+\frac{1}{\lambda _{I,a}}\right)\frac{ \Gamma \left(s-1/2\right)}{  \Gamma (s)} \zeta (2 s-1)+ \frac{\pi}{ 2 \lambda _{I,a} \lambda _{I,b}} \frac{\Gamma (s-1)}{ \Gamma (s)} \zeta (2 s-2) \nn \\
&& \hspace{3 cm}  -\frac{2\sqrt{\pi}}{\lambda_{I,a} \Gamma(s)} \sum_{n,l=1}^{\infty} \left(\frac{l}{n\pi/\lambda_{I,a}}\right)^{ -s+1/2} K_{ -s+1/2}\left( \frac{ 2  n l \pi  }{\lambda_{I,a}} \right) \nn \\
&& \hspace{3 cm} -\frac{2\sqrt{\pi}}{\lambda_{I,b} \Gamma(s)} \sum_{n,l=1}^{\infty} \left(\frac{l}{n\pi/\lambda_{I,b}}\right)^{ -s+1/2} K_{ -s+1/2}\left( \frac{ 2  n l \pi  }{\lambda_{I,b}} \right) \nn \\
&& \hspace{3 cm} +\frac{2 \pi} {\lambda_{I,a}\lambda_{I,b} \Gamma(s)} \sum_{n,l=1}^{\infty} \left(\frac{l}{n\pi/\lambda_{I,a}}\right)^{-s+1} K_{-s+1}\left( \frac{ 2  n l \pi  }{\lambda_{I,a}} \right) \nn \\
&& \hspace{3 cm} +\frac{2 \pi} {\lambda_{I,a}\lambda_{I,b} \Gamma(s)} \sum_{n,l=1}^{\infty} \left(\frac{l}{n\pi/\lambda_{I,b}}\right)^{-s+1} K_{-s+1}\left( \frac{ 2  n l \pi  }{\lambda_{I,b}} \right) \nn \\
&&  \hspace{2 cm} +\frac{4 \pi} {\lambda_{I,a}\lambda_{I,b} \Gamma(s)} \sum_{n,m,l=1}^{\infty} \left(\frac{l }{\sqrt{\left(n\pi/\lambda_{I,a} \right)^2+\left(m\pi/\lambda_{I,b} \right)^2 }}\right)^{-s+1 }\nn \\
&&  \hspace{6 cm} \times K_{-s+1}\left( 2  l \sqrt{\left(n\pi/\lambda_{I,a} \right)^2+\left(m\pi/\lambda_{I,b} \right)^2 } \right)
\label{eq-10}
\eea
Finally, performing the limit $s \rightarrow 0$ by noting that
\bea
\lim_{s \rightarrow 0} \frac{g(s)}{\Gamma(s)}=g(0), \nn
\eea
 the zero-point energy \eqref{eq-5} turns to
\bea
&&E_0(a,b,T) =-\frac{k_B T}{2} \sum_I \Bigg(\frac{\zeta(3)}{4\pi \lambda _{I,a} \lambda _{I,b}}-\frac{\pi }{12 \lambda _{I,a}}-\frac{\pi }{12 \lambda _{I,b}} \nn \\
&&\hspace{6 cm}+\frac{\pi}{ 12} \frac{ \lambda _{I,a}}{ \lambda _{I,b}}-\frac{\ln\lambda _{I,a}}{2} 
 +\sum_{n,m=1}^{\infty} \frac{\exp{\left(- 2  m n\pi \lambda _{I,a}/\lambda _{I,b} \right)}}{m}\nn \\
&&\hspace{6 cm} -\sum_{n,m=1}^{\infty} \frac{\exp{\left(- 2  m n\pi /\lambda _{I,a} \right)}+\exp{\left(- 2  m n\pi /\lambda _{I,b} \right)}}{n} \nn \\
&&\hspace{6 cm} +2 \sum_{n,m=1}^{\infty} \frac{m}{n} \left[\frac{K_1(2 n m\pi/\lambda_ {I,a})}{\lambda_ {I,b}}+\frac{K_1(2 n m\pi/\lambda_ {I,b})}{\lambda_{I,a}} \right]\nn \\
&&\hspace{6 cm} +4 \sum_{n,m,l=1}^{\infty} l \frac{K_1\left( 2  l  \pi \sqrt{ n^2/\lambda_{I,a}^2+ m^2/\lambda_{I,b}^2 } \right)}{ \sqrt{ n^2 \lambda_{I,b}^2+ m^2 \lambda_{I,a}^2 }} \Bigg)  \label{eq-11}
\eea
Note that the first three terms of the above equation are the contributions of unbounded system (i.e. a rectangle with infinite side lengths) which result in nonzero terms for the Casimir force at the limits $a,b \rightarrow \infty$, note Eq. \eqref{eq-4}.  Hence these terms should be subtracted, to  find the phononic Casimir energy  of the membrane
\bea
&&E_C(a,b,T) =\frac{k_B T}{2}\sum_I \Bigg(\frac{\ln\lambda _{I,a}}{2} -\frac{\pi}{ 12} \frac{ \lambda _{I,a}}{ \lambda _{I,b}}
 -\sum_{n,m=1}^{\infty} \frac{\exp{\left(- 2  m n\pi \lambda _{I,a}/\lambda _{I,b} \right)}}{m}\nn \\
&&\hspace{4 cm} +\sum_{n,m=1}^{\infty} \frac{\exp{\left(- 2  m n\pi /\lambda _{I,a} \right)}+\exp{\left(- 2  m n\pi /\lambda _{I,b} \right)}}{n} \nn \\
&&\hspace{4 cm} -2 \sum_{n,m=1}^{\infty} \frac{m}{n} \left[\frac{K_1(2 n m\pi/\lambda_ {I,a})}{\lambda_ {I,b}}+\frac{K_1(2 n m\pi/\lambda_ {I,b})}{\lambda_{I,a}} \right]\nn \\
&&\hspace{4 cm} -4 \sum_{n,m,l=1}^{\infty} l \frac{K_1\left( 2  l  \pi \sqrt{ n^2/\lambda_{I,a}^2+ m^2/\lambda_{I,b}^2 } \right)}{ \sqrt{ n^2 \lambda_{I,b}^2+ m^2 \lambda_{I,a}^2 }} \Bigg)  \label{eq-12}
\eea
Note that in contrast to the heat kernel approach, our technique has resulted in an \emph{ exact } equation for the Casimir energy.  In fact an exact term such as the third term of the above equation, which would turn out to be an important term,  can not be simply obtained through the heat kernel approach. For sufficiently large temperatures (with respect to the effective temperature $\theta_{\textrm{LA},b}$), the last three terms  can be neglected  to find
\bea
E_C(a,b,T)\approx  k_B T \Bigg(\frac{1}{4}\sum_I\ln \left[\frac{\hbar \upsilon_I}{ 2a k_B T} \right] -\frac{\pi}{ 12} \frac{ b}{a}
 -\sum_{n,m=1}^{\infty} \frac{\exp{\left(- 2  m n\pi b/a \right)}}{  m} \Bigg)    \label{eq-13}
\eea
Then the phononic Casimir forces  can be defined as
\bea
&&F_{\textrm{C,a}} (a,b,T)  \equiv- \frac{\partial  E_C }{\partial a} \approx \frac{ k_B T}{a} \left(\frac{1}{2} -\frac{\pi}{ 12} \frac{ b}{a }+\frac{ b}{a } S\left(\frac{b}{a}\right) \right) \nn \\ 
&&F_{\textrm{C,b}} (a,b,T)  \equiv - \frac{\partial  E_C }{\partial b} \approx \frac{ k_B T}{b} \left(\frac{\pi}{ 12} \frac{ b}{a}
 -\frac{ b}{a } S\left(\frac{b}{a}\right) \right)
 \label{eq-14} 
\eea
in which 
\bea
S(x) \equiv  \sum_{n,m=1}^{\infty} 2 n\pi\exp{\left(- 2  m n\pi x \right)}; \ \  x \leq 1
\nn 
\eea
 Through numerical computations, we have $S(x)\geq S(1) \approx 0.01$, hence the absolute value of $F_{\textrm{C,a}}$ as well as $F_{\textrm{C,b}}$ increases by decreasing the ratio $b/a$. However $F_{\textrm{C},a}$  is always positive-valued (i.e. repulsive), while $F_{\textrm{C},b}$ is positive-valued for $b/a\gtrapprox 0.5$, and  negative-valued (i.e. attractive) for $b/a\lessapprox 0.5$. Note that for a square trench ($a=b$) we have
\bea
F_{\textrm{C,a}}  = F_{\textrm{C,b}}  \approx 0.25 \frac{ k_B T}{a} 
 \label{eq-15} 
\eea
so for a square membrane the phononic Casimir force is always repulsive. 

Note that an asymptotically appropriate expression for sufficiently small  temperatures (in comparison to $\theta_{\textrm{TA},a}$), could be obtained by applying the Poisson sum \eqref{eq-6} to the $l$-sum  in Eq. \eqref{eq-5}. Then  after some similar calculations one would find
\bea
&&E_{\textrm{C}}(a,b)=\left(\frac{\pi}{48} -\frac{\zeta(3)}{16\pi} \frac{b}{a} -  \sum_{n,m=1}^\infty \frac{n}{2m} K_1\left(2\pi n m \frac{b}{a}\right) \right) \frac{\hbar}{a} \sum_I \upsilon_I \nn \\
&& \hspace{3 cm}- \frac{k_B T}{2} \sum_I \Bigg(\frac{\pi }{12 \lambda _{I,a}}+\frac{\pi }{12 \lambda _{I,b}}-\frac{ \zeta(3)}{4\pi \lambda _{I,a} \lambda _{I,b}} \nn \\ 
&&\hspace{5 cm} + \sum_{n,m,l=1}^{\infty} \frac{\exp\left(- 2  l  \pi \sqrt{ n^2 \lambda_{I,a}^2+ m^2 \lambda_{I,b}^2 } \right)}{l}\Bigg)
\label{eq-16}
\eea
which, for sufficiently small temperatures, can be approximated by its first line.

\section{Application to Graphene resonator} \label{sec-3}

 As we know, the dynamics of a sufficiently thin membrane being perfectly flexible, with a sufficiently large pre-tension, can be described effectively by  surface   harmonic oscillation modes \cite{VPE}.  This is a valid consideration  e.g. for a suspended monolayer  Graphene sheet with a rather large initial tension arisen   from the strong  VdW adhesion which clamp the Graphene sheet to the sidewalls of the trench \cite{BT,GV1,GV2,GV3,GV4,GT1,GT2,GT3}.  Note however that this may  be not a good approximation for thick multilayer Graphene membranes, which have larger  values of the bending rigidity being not negligible compared to the built-in tension of the Graphene \cite{BT}. 
 
Now to find an appropriate physical interpretation for the phononic Casimir forces for  such a fully clamped membrane with a sufficiently large pre-tension, we introduce the total ground state energy ($E_\textrm{g}$) of the  membrane as the sum of the (classical) tensional energy and the phononic Casimir energy. Then the change in $E_\textrm{g}$, of a fully clamped membrane with the  pre-tension $\tau_0$, due to infinitesimal changes $\delta a$ and  $\delta b$,  can be written as
\bea
\delta E_\textrm{g} &=&(b\, \tau_0)\delta a+ \frac{\partial E_{\textrm{C}}}{\partial a}\delta a + (a\, \tau_0)\delta b+ \frac{\partial E_{\textrm{C}}}{\partial b}\delta b
\nonumber \\ &=& b\big(\tau_0 - \frac{1}{b} F_{\textrm{C,a}}\big)\delta a + a\big(\tau_0 - \frac{1}{a} F_{\textrm{C,b}}\big)\delta b. \label{eq-17}
\eea
 Hence one can take the quantum corrected tensions of the membrane as
\bea
\tau_a=\tau_0 - \frac{1}{b} F_{\textrm{C,a}} \nn \\
\tau_b=\tau_0 - \frac{1}{a} F_{\textrm{C,b}} \label{eq-18}
\eea
i.e. the phononic Casimir forces contribute actually  as quantum corrections to the  pre-tension of the membrane. For sufficiently large temperatures the Casimir forces for the fully clamped membrane  are given by Eq. \eqref{eq-14}, so the corrected pre-tensions of the membrane over a rectangular trench can be written as
\bea
\tau_{a,b} (a,b,T) \approx \tau_0 + \Delta_{a,b} (a,b,T)   \label{eq-19}
\eea
 in which
 \bea
&& \Delta_a (a,b,T)\equiv - \frac{ k_B T}{b^2}  \left[   \frac{1}{2}\frac{b}{a}-\frac{\pi}{ 12} \left(\frac{b}{a} \right)^2+ \left(\frac{b}{a} \right)^2 S \left(\frac{b}{a}\right) \right]    \nn \\
&&  \Delta_b(a,b,T) \equiv  \frac{ k_B T}{b^2} \left[- \frac{\pi}{ 12} \left(\frac{b}{a} \right)^2+ \left(\frac{b}{a} \right)^2 S \left(\frac{b}{a}\right) \right]   \label{eq-20}
\eea
with  $S(x)$ as before. Note that  for a square trench ($a=b$),
\bea
 \tau_a=\tau_b   \approx \tau_0 -  0.25 \frac{ k_B T}{a^2}. \label{eq-21}
\eea
while    for a narrow trench $b/a \ll 1$ we have
\bea
&&\tau_{a,b} (a,b,T) \approx \tau_0 \mp   \frac{ k_B T}{a^2} S \left(\frac{b}{a}\right)     \label{eq-22}
\eea
in which the signs ``$-$" and ``$+$"  are referred to the indexes ``$a$"  and ``$b$"  respectively. According to previous numerical computations (see below Eq. \eqref{eq-14}), for a rectangular trench, the phononic Casimir corrections always decrease $\tau_a$ , while $\tau_b$ decreases for $b/a\gtrapprox 0.5$, and  increases for $b/a\lessapprox 0.5$.  So as a result  the phononic Casimir corrections generally break the tensional isotropy of the fully clamped membrane (i.e. $\tau_a \neq \tau_b$). However for a square trench, the phononic Casimir force always decreases the membrane pre-tension.

 \begin{figure}[b]
\centering
\includegraphics[width=8 cm, height=5 cm]{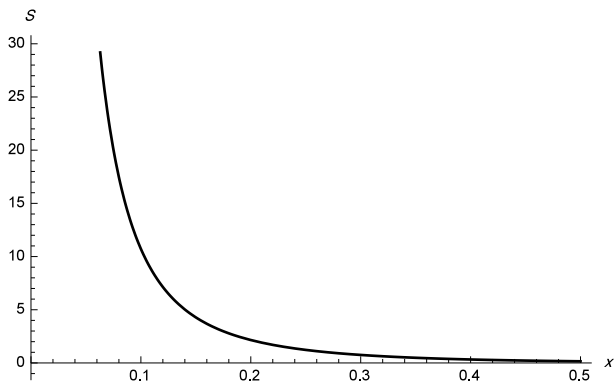}
\caption{Numerical plot of  $S(x)$}
\label{S}
\end{figure}

Now we can use experimental values to obtain numerical results for the above corrections. As we previously mentioned, a rather large tension is induced by the strong Van der Waals adhesion between the Graphene and the sidewalls of the trench.  The (surface) pre-tension for suspended   Graphene sheets,  at room temperature has been estimated as being of the order of $10^{-3}$ - $10^{-2} \textrm{N/m}$, see e.g. \cite{GT1,GT2,GT3}. The one order of magnitude difference in these experimental values, has been attributed to e.g. the external forces applied during the Graphene fabrication, and/or the influences of the Graphene adsorbates \cite{GT2,GTT1}. So  the value of the VdW-induced tension of the Graphene, can be taken as  $ \sim 10^{-3} \textrm{N/m}$.   Having the acoustic wave velocities  of the  Graphene as  $\upsilon_{\textrm{TA}} \approx 13.6 \textrm{ km/s} $ and  $\upsilon_{\textrm{LA}} \approx 21.3 \textrm{ km/s}$ (see e.g. \cite{GP}), the (largest) effective temperature  of a Graphene sheet suspended over a rectangular trench with the side-lengths both  in the order of few micrometers,  would be obtained as  $\theta_{\textrm{LA},b}\sim 0.1 \textrm{K}$, see Eq. \eqref{eq-4} (having $\hbar \approx 1.05 \times 10^{-34} m^2 kg/s$ and $k_B \approx 1.38 \times 10^{-23} m^2 kg/s^2 K$) . This effective temperature  is obviously far smaller than   ordinary temperatures, hence the  asymptotic expressions for the Casimir energy \eqref{eq-13}, Casimir forces \eqref{eq-14}, and the corrected tensions \eqref{eq-19},  are valid with a high degree of accuracy, for a wide range of   accessible temperatures. 

As one can see from Eq. \eqref{eq-19}, at room temperature ($ \approx 300 \textrm{K}$), the  temperature corrections $\Delta_{a,b}$  for a trench with side-lengths both in the order of few micrometers,  would be obtained as $\sim 10^{-10} \textrm{N/m}$, which, in comparison to the  built-in  tension $ \tau_0$ ($\sim  10^{-3} \textrm{ N/m}$), is completely negligible. However,  the absolute-value of the temperature correction, $|\Delta_{a,b}|$,  increases by decreasing the side-lengths $a,b$ as well as  by decreasing the ratio $b/a$, since as one can see in the numerical  plot \ref{S}, $S(x)$  increases extremely by reducing ``$x$" . As a result,   for  ``$b$" and ``$a$" e.g.  in the order of few  nanometers and few micrometers, respectively,   using Eq. \eqref{eq-22} with  $S\left(0.001 \right) \approx 1.3 \times 10^{5}$, one can find a value   of the order of $10^{-4} \textrm{N/m}$ for    $|\Delta_{a,b}|$ at room temperature, which is noticeable in comparison to the  built-in  tension $\tau_0$. Note that for $b \approx 10^{-9} \textrm{m}$, the effective temperature $\theta_{\textrm{LA},b}$ would be in  the order of $100 \textrm{ K}$, which is still sufficiently smaller than the room temperatures, so that the asymptotic expressions \eqref{eq-13}, \eqref{eq-14} and \eqref{eq-19} are still valid with a good degree   of accuracy.  These temperature corrections  can have even more importance for Graphene membranes   at  larger temperature,  since actually the built-in tension of the Graphene sheet decreases by increasing the temperature, see Refs. \cite{GTT1,GTT2,GTT3}, while $|\Delta_{a,b}|$ increases by increasing the temperature. Then the change in the Graphene tensions ($\Delta \tau_{a,b}$), in terms of the temperature change ($\Delta T$), can be given as
\bea
\Delta \tau_{a,b} \approx \mp   \frac{ k_B \Delta T}{a^2} S \left(\frac{b}{a}\right) \label{eq-23}
\eea
These temperature corrections can be applied for adjusting the built-in tension of the Graphene resonator, specifically at higher temperatures. But as we know, any change in the Graphene tension would be observable as a change in its resonance frequencies, specifically the   fundamental resonance frequency, see e.g. \cite{GT1}. However, as a result of anisotropic tension of the Graphene due to the acoustic Casimir corrections, the fundamental resonance frequency of the Graphene membrane \cite{BT}, should be corrected as
\be
f_{11} =   \sqrt{ \frac{1}{2\mu}\left(\frac{\tau_a}{a^2} +\frac{\tau_b}{b^2} \right)}  \label{eq-24}
\ee
in which, $\mu \approx 10^{-6} \textrm{kg/m}^2$ is the surface mass-density of the Graphene (see e.g. \cite{BT,GT1,GT2}), and $\tau_{a,b}$ are given by Eq. \eqref{eq-19}. Then for $b/a \ll 1$,  using Eq. \eqref{eq-23}, the change  (in  the squared value)   of the fundamental resonance frequency  \eqref{eq-24}, in terms of the temperature change, would be given as
\bea
 \Delta f_{11}^2 & =&   \frac{\Delta\tau_a}{a^2} +\frac{\Delta \tau_b}{b^2} \nn \\
&\approx&  \frac{ k_B \Delta T}{2\mu a^2 b^2} S \left(\frac{b}{a}\right).
  \label{eq-25}
\eea

\section{Conclusion and  remarks}

  We have applied  a  useful technique (see Eqs. \eqref{eq-5}-\eqref{eq-11}) to obtain the  Casimir energy at finite temperature for acoustic phonons in a fully clamped  sufficiently tensioned membrane suspended over a rectangular trench, see Eq. \eqref{eq-12}, and  the corresponding  Casimir forces \eqref{eq-14} have been interpreted as quantum temperature-dependent corrections to the  built-in  (surface) tensions of the membrane, see Eqs. \eqref{eq-18}-\eqref{eq-20}. We have shown that these temperature corrections  generally break the tensional isotropy of a  membrane over a rectangular trench, and can decrease or increase the  pre-tensions of the membrane, while for a square trench, the corrections always decrease the membrane pre-tension, see the numerical discussions below Eqs. \eqref{eq-14} and \eqref{eq-22}.  

  We have obtained numerical results for  a monolayer Graphene sheet, using the experimental values  given  in Refs. \cite{GT1,GT2,GT3}, and  have demonstrated  that for a narrow rectangular trench with side-lengths in the order of few   nanometers and  few  micrometers,   the temperature corrections to the pre-tensions, at room temperature, would be of the order of $10^{-4} N/m$, which is expected to be noticeable in comparison to the VdW-induced built-in tension of the Graphene sheet, see the numerical discussion above Eq. \eqref{eq-23}.    Consequently we have introduced a corrected version for the fundamental resonance frequency of the suspended Graphene membrane, and have obtained its change in terms of  the temperature change, see Eq. \eqref{eq-25}. 

  These  temperature corrections would find even more importance  for Graphene membranes  at larger temperature, since the built-in tension of the Graphene decreases by increasing the temperature   \cite{GTT1,GTT2,GTT3}, while  (the absolute value of)  the temperature correction increases by increasing the temperature, see Eq. \eqref{eq-23}.  Hence, these temperature corrections can be applied for adjusting  the built-in tension  of the Graphene resonator, specifically at higher temperatures.
 
\subsection*{ Acknowledgment}
We thank  Sayed Akbar Jafari for his valuable comments, and Saeed  Qolibikloo for his helps during the numerical computations.

\end{document}